\documentclass[a4paper, twocolumn, superscriptaddress,prb]{revtex4}  
\usepackage{amsmath,amssymb}
\usepackage{graphicx}
\usepackage{textcomp}

\begin{document}

\title{\huge Shot noise limited characterization of femtosecond light pulses}

\author{O. Schwartz}
\email{osip.schwarz@weizmann.ac.il}
\author{O. Raz}
\author{O. Katz}
\author{N. Dudovich}
\author{D. Oron}
\affiliation{Department of Physics of Complex Systems, Weizmann
Institute of Science, Rehovot, Israel}

\maketitle
\textbf{Probing the evolution of physical systems at the femto- or
attosecond timescale with light requires accurate characterization
of ultrashort optical pulses. The time profiles of such pulses are
usually retrieved by methods utilizing optical nonlinearities, which
require significant signal powers and operate in a limited spectral
range\cite{Trebino_Review_of_Scientific_Instruments97,Walmsley_Review_09}.
We present a linear self-referencing characterization technique
based on time domain localization of the pulse spectral components,
operated in the single-photon regime. Accurate timing of the
spectral slices is achieved with standard single photon detectors,
rendering the technique applicable in any spectral range from near
infrared to deep UV. Using detection electronics with about $70$ ps
response, we retrieve the temporal profile of a picowatt pulse train
with $\sim10$ fs resolution, setting a new scale of sensitivity in
ultrashort pulse characterization.}

Complete characterization of an optical pulse implies the knowledge
of its time profile, or, equivalently, its spectrum and the spectral
phase. Despite the rapid progress of detection electronics
\cite{Real_time_optical_waveform_measurement_NatPhot2010}, direct
measurement of a subpicosecond time profile remains a difficult
task. The spectral phase can be reconstructed from the interference
with a known coherent reference pulse\cite{Phasometer_Dan_JOSAB97}.
Usually, however, such reference is unavailable, and the pulse shape
retrieval requires a self-referencing technique.

Self-referenced characterization of an ultrashort pulse is usually performed by
interacting it with its replica in a nonlinear optical medium. During the last
two decades nonlinear optical characterization techniques such as
FROG\cite{Trebino_FROG_JOSAA93} and SPIDER\cite{SPIDER_OptLett98} have become a
standard tool in ultrafast research. Employing optical nonlinearities, however,
imposes substantial limitations on power, spectral range and complexity of the
characterized pulse\cite{Trebino_book_2002}.

These difficulties stimulated attempts to develop spectral phase
retrieval methods using only linear optics and ``slow'' detectors.
The ability to characterize pulses with a temporal resolution
significantly better than the detector response is intimately
related, as in precision spectroscopy, to the system noise
characteristics. Thus, while there are several approaches that allow
for reliable picosecond waveform reconstruction
\cite{Walmsley_Review_09,
Linear_characterization_review_Dorrer_JOSAB2008,
Linear_Spectrogrrams_Reid_Harvey_2007IPTL,
Sonogram_megahertz_genrator_IM95}, few works have demonstrated
linear optical characterization with femtosecond resolution, whether
by spectral
shearing\cite{ElectroOptic_Modulator_200fs_Kang_Dorrer_OL2003}, time
resolved\cite{Kockaert_StreakCamera_JQE04} interferometry, or
direct\cite{Sonogram_reflectometry_Moon_PhotTecLett2009} or cross
correlation\cite{Prein_integrated_detector_OC96} sonogram
measurements. In all these, the extracted quantity is the derivative
of the spectral phase $\phi(\omega)$ with respect to frequency
$\omega$. This can be thought of as the arrival time of a spectral
slice of a pulse, provided that the slice is narrow enough so that
the phase within it can be approximated by a linear
function\cite{Chilla_Martinez_SHG_Gating_OL91}. The best reported
temporal resolution in linear chracterization was of $50$fs at a
power level of $0.1$mwatt \cite{Prein_integrated_detector_OC96}. All
linear methods significantly fall behind the few nanowatt
sensitivity recently demonstrated in nonlinear
characterization\cite{Ultralow_Power_Frog_Poled_Crystal_JOSAB08}.
Yet, as we show below, this is not a true limitation of linear
characterization.

In this Letter we demonstrate a linear self-referencing ultrashort
pulse characterization technique with shot noise limited temporal
resolution. The technique relies on precise timing of spectral
slices using photon counting instrumentation. We take advantage of
the fact that, depending on the signal to noise ratio, the timing
error can be much smaller than the detector response time. For a
train of identical pulses, the signal to noise ratio can be improved
by averaging over many pulses. The overall resolution then increases
as $\sqrt{N}$, where $N$ is the number of photons detected (shot
noise resolution scaling). Utilizing a differential measurement
scheme with the signal and the reference pulses coupled into the
same detector, we demonstrate the $\sqrt{N}$ uncertainty scaling
down to few femtoseconds. The principle of converting timing
accuracy into temporal resolution  is analogous to superresolution
microscopy techniques relying on spatial localization of individual
fluorophores \cite{Nanometer_Localization_Thompson02BJ,
PALM_Betzig_Science2006, STORM_NatMet06,
SubNanometer_Localization_Chu_Nature2010}.

\begin{figure}
\begin{picture}(250,430)
\put(0,180){\includegraphics[natwidth=720,natheight=720,clip=true,scale=0.35]{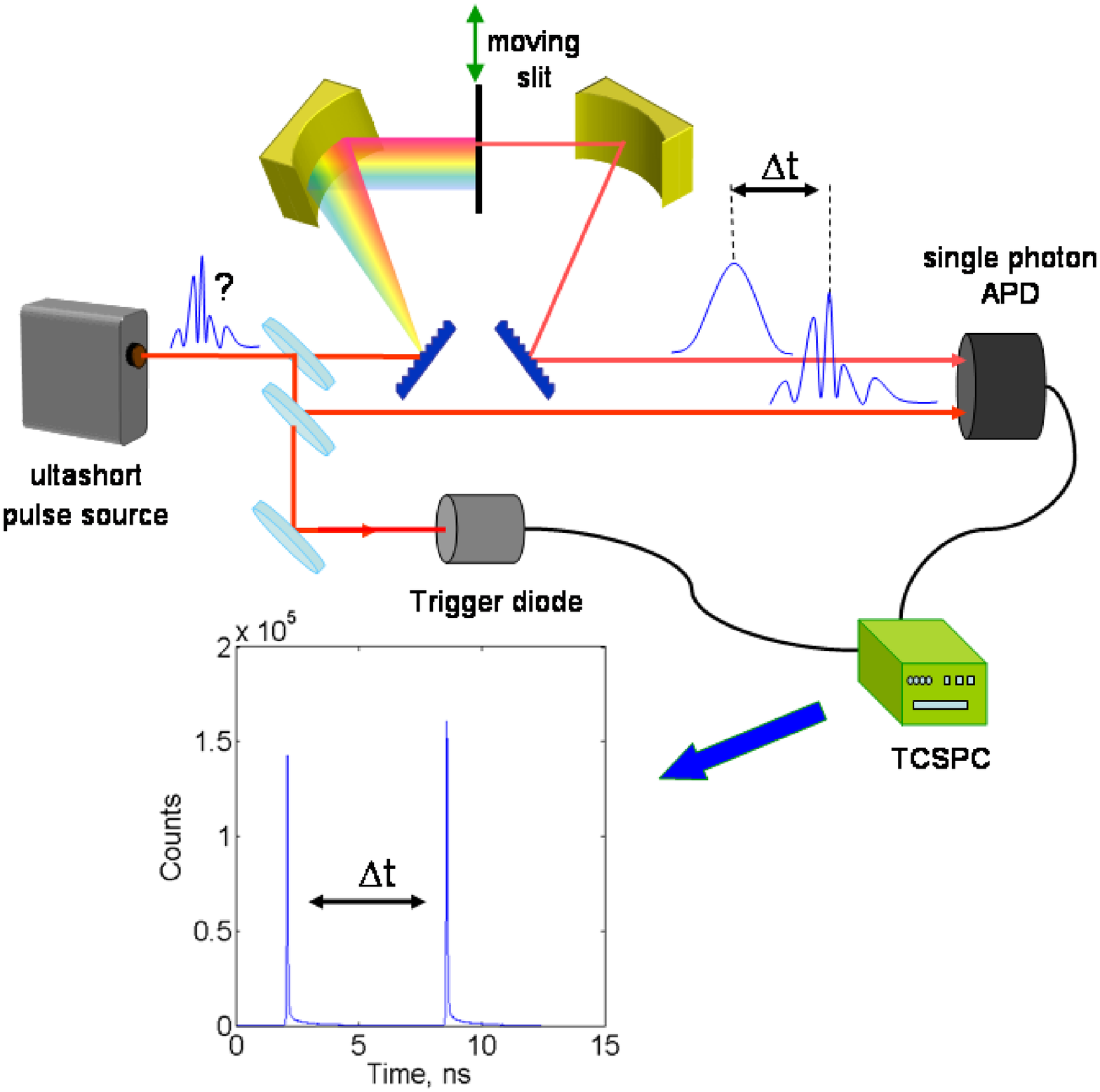}}
\put(0,0){\includegraphics[width=85mm]{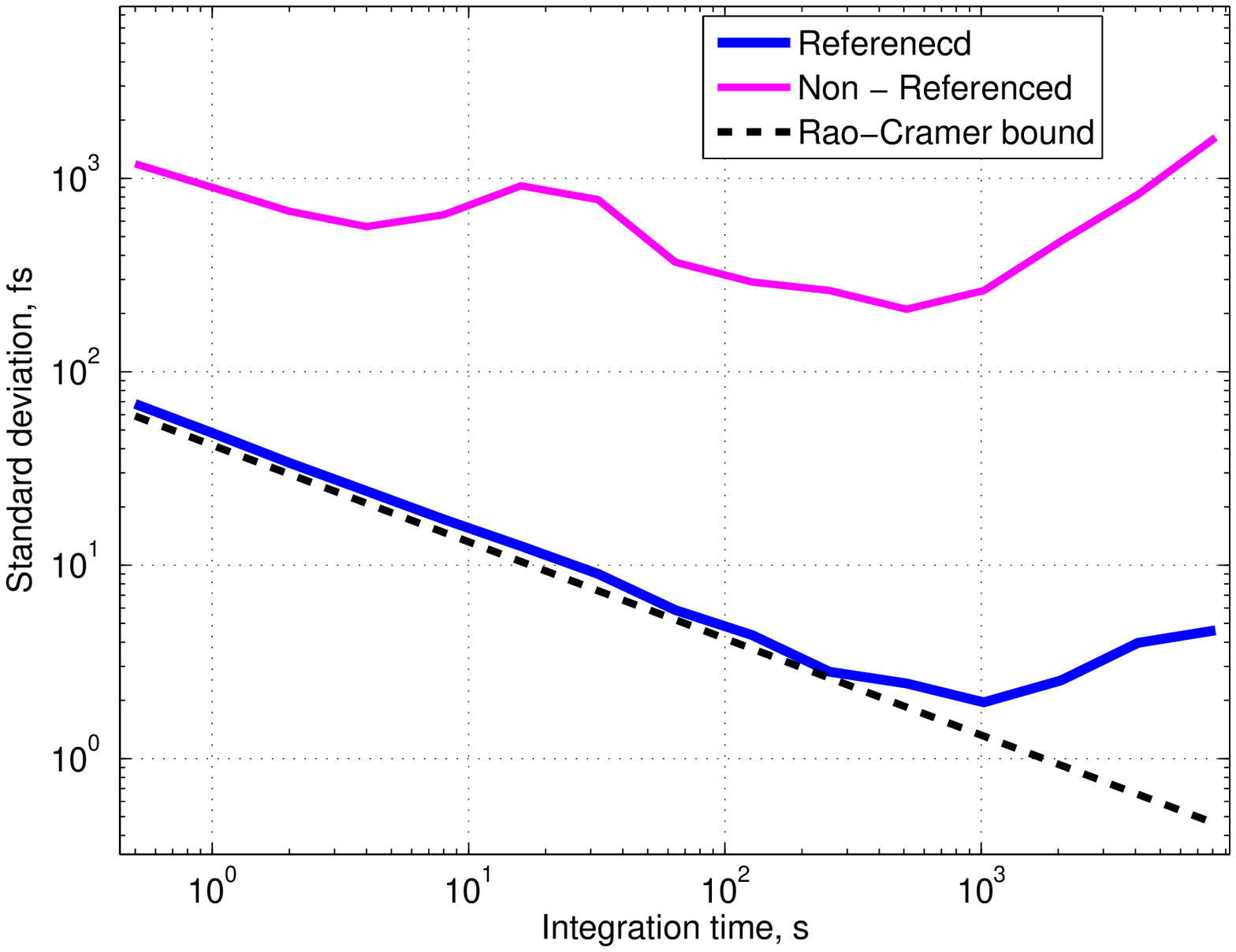}}
\put(0,400){\mbox{\Large \textbf{a}}} \put(0,150){\mbox{\Large
\textbf{b}}}
\end{picture} \caption{\textbf{a) Experimental setup.} The characterized pulse is passed
through a 4f pulse shaper selecting a spectral component, and is
detected by a single photon counting APD. One replica of the
original pulse is used as a trigger signal for the TCSPC module,
while the other, serving as an additional reference, is directed
into the signal detector. The inset shows a sample histogram with
the two peaks corresponding to the signal and the reference pulses.
\textbf{b) Allan variance.} The plot shows the standard deviation of
the measured pulse arrival time as a function of the integration
time. Two replicas of the same pulse separated by a constant delay
were used as the signal and the reference, each attenuated to $1$
MHz detection rate. The arrival time of the signal pulse was
measured by finding the absolute position of the peak in the
histogram (magenta) and by finding its position relative to the
reference pulse (blue). The dashed black line represents the
Cramer-Rao minimum variance for this measurement (see Supplementary
Sections 1 and 2)}\label{setup}
\end{figure}

We implemented this method using a fast single photon avalanche
photodiode (idQuantique id100) for signal detection and a time
correlated single photon counting (TCSPC) module (PicoHarp 300) for
electric pulse timing. The ability of TCSPC electronics to detect
delays as large as few nanoseconds allowed for very large temporal
dynamic range. The layout of our setup is presented in
Fig.~\ref{setup}. The FWHM instrument response was measured to be
about $70$ ps. Spectral slices were selected by a 4f pulse
shaper\cite{Weiner_RSI2000} with a mobile slit in the Fourier plane.
At every slit position, the data was collected for $0.5$s; the
spectrum was scanned several times to achieve the desired
integration time. For every slice, we generated a histogram of
arrival times of the signal relative to the trigger pulse. The
arrival time was obtained by finding the centroid of the histogram
peak (see Supplementary Section 1 for details of the data
treatment).

The femtosecond relative delays between the spectral components are
much smaller than the systematic shifts induced by the intensity
dependence of the detector response and the drifts of the centroid
due to temperature fluctuations. To overcome this difficulty, we
introduced another reference beam: a delayed replica of the original
pulse was directed into the detector along with the signal beam.
Since the photon flux was lower than the saturation threshold, the
detector was randomly activated by photons coming from one of the
two beams. The arrival time of each spectral component was then
given by the relative delay of the two corresponding peaks in the
histogram. An Allan variance plot, demonstrating the measurement
timing error as a function of the integration time for a single peak
and for the interval between the two peaks is shown in
Fig.\ref{setup} b. While the unreferenced measurement deviates from
the $\sqrt{N}$ scaling by an order of magnitude already at $1$ s
integration time, the relative delay measurement remains shot noise
limited up to $\sim1000$ seconds of integration, corresponding to a
sub-cycle standard deviation of $2$ fs.

We first tested our setup by characterizing test pulses from a Ti:Sapphire
mode-locked oscillator. We retrieved the 85 fs transform limited laser output,
as well as a chirped pulse obtained by passing it through a 6" slab of F3
glass. For comparison, the same pulses were characterized by a second harmonic
FROG setup utilizing a $100$ $\mu$m BBO crystal. For the linear measurement,
the characterized beam was attenuated to about $1$ photon per pulse
(approximately $10$ pW), leading to $\sim 10^6$ detector counts per second. The
integration time was $40$ s per data point. Prior to the measurements, a
calibration curve accounting for any uncompensated dispersion in the setup was
recorded using a transform limited pulse.

\begin{figure}
\includegraphics[width=90 mm]{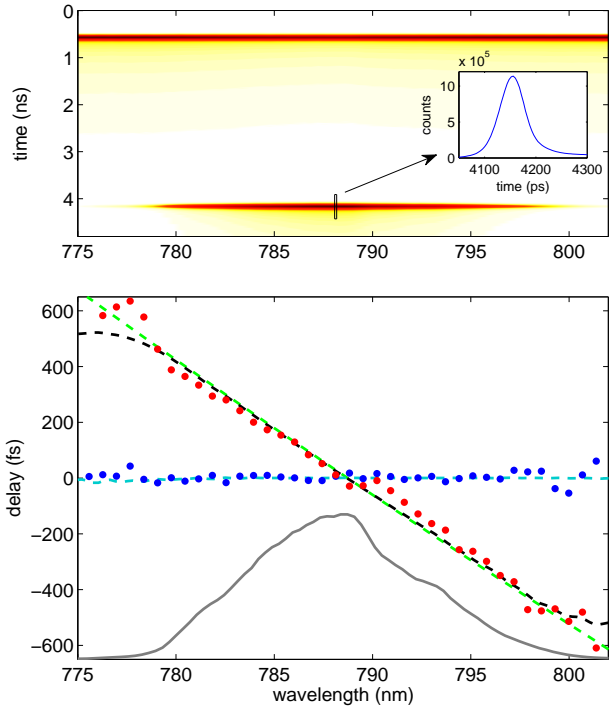}%
\caption{\textbf{Characterization of a 85 fs transform limited pulse
and a pulse chirped by a 6" long F3 glass slab.} \textbf{a)} Density
plot of a typical raw measurement trace featuring two peaks, the
reference and the signal. The inset shows a cross section of the
signal peak with FWHM of about $70$ ps. \textbf{b)} Measured delays
of the spectral components for the transform limited and the chirped
pulse (represented by the blue and red dots, respectively). The
dashed black and cyan lines show the delays derived from the FROG
measurements, and the dashed green line represents Sellmeier
equation. The gray solid curve in the lower part of the plot shows
the spectrum of the pulses in arbitrary units.} \label{glass}
\end{figure}

The measured spectral component delays with and without the glass slab are
presented in Fig.~\ref{glass}. The resolution of the measurements, estimated by
calculating the standard deviation in the range $784$ to $789$ nm after
subtraction of the linear trend, was $9.3$ fs. for the glass measurement and $10.6$ fs for the transform limited pulse.
These values are in good agreement with the $\sqrt{N}$ scaling of
resolution discussed above (see also Supplementary Section 1). As
expected, the plots show linear delay dependence on wavelength, with
the slope of $-41.14$fs/nm. The 10\% systematic deviation from the
results of the FROG measurements ($-46.4$fs/nm) and from Sellmeier
equation for the F3 glass\cite{Schott} ($-47.4$fs/nm), which are
also shown in Fig.~\ref{glass} can be attributed to a dependence of
the calibration curve on the exact alignment of the beam into the
detector. This coupling sensitivity is probably due to the fact that
in our detector the entrance aperture is connected to the actual APD
by a short length of a multimode fibre. The temporal pulse profiles
reconstructed from our measurement and from the FROG data are shown
in Supplementary Fig.2.
\begin{figure}
\includegraphics[width=90 mm]
{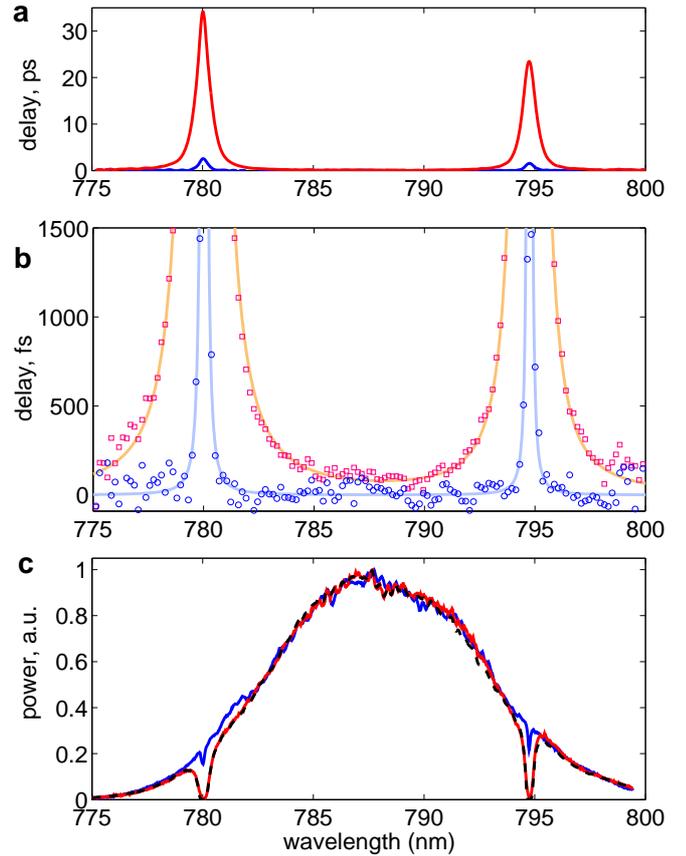}%
\caption{\textbf{Measurement of the spectral delay of pulses passed through a
Rb cell.} \textbf{a)} Measured delays of spectral components at
100\textcelsius\ (blue) and 210\textcelsius\ (red). The integration time was
$4$s per point at 100\textcelsius\ and $8$s per point at 210\textcelsius.
\textbf{b)} A magnified view of the same (markers) fitted with a resonance
delay profile (lines). \textbf{c)} Normalized spectrum of the transmitted pulse
at 100\textcelsius\ (blue) and 210\textcelsius\ (red) measured at a resolution
of $0.05$ nm. (The dashed black line shows the resonant absorbtion profile
corresponding to the fit of 210\textcelsius\ spectral delay data.)}
\label{rubidium}
\end{figure}

The system's ability to reconstruct complicated pulses was tested by
passing the beam through a hot Rb vapor cell. Rubidium has two
narrow absorption lines in the relevant spectral range, at 780 and
795 nm. At large optical densities, achieved by heating the cell,
the transmitted pulses are distorted by the resonant lines
dispersion profile. The resulting complex pulses contain temporally
extended tails, which make their nonlinear optical characterization
challenging. Each resonance modifies the electric field of the pulse
multiplying it by a factor of $f(\omega) =\exp\left[ - \alpha l
\;\left(1\left/(1 + i\omega T_2)\right.\right)\right]$, where $T_2$
is the inverse pressure broadened linewidth, $\alpha l$ is the
optical density and $\omega$ is the frequency detuning
\cite{Dan_Nirit_Rubidium_PRL02}. The corresponding spectral slice
delay is then $t(\omega) = \alpha l T_2 \frac{1- \omega^2
T_2^2}{\left( 1 + \omega^2 T_2^2\right)^2}$. The results of the
measurements are presented in Fig. \ref{rubidium}. The spectral
resolution of the system was set to $0.5$ nm, thus the observed
relative delays were weighted averages of the true delays over that
range. The spectral delay measurements were complemented by a higher
resolution absorbtion spectrum measurements shown in
Fig\ref{rubidium}c. At room temperature, no effect is observed. At
100\textcelsius\ the delay peaks around the two resonances are
clearly visible, although the shape of the delay peaks is dominated
by the instrument response. As the temperature of the cell is
increased to $210$\textcelsius, the delay in the Lorentzian tails of
the above profile becomes significant and is clearly observed.
According to the equations above, the magnitude of the tails of the
spectral delay curve is $\alpha L \left(T_2 \omega^2\right)^{-1} $,
while the magnitude of the Lorentzian tails of absorbtion lines in
the transmitted spectrum (shown in Fig \ref{rubidium}c) can be
written as $2 \alpha L \left(T_2 \omega\right)^{-2}$. Extracting the
two coefficients for the $780$ nm resonance from the
$210$\textcelsius\ data (see Supplementary Section 4) and comparing
them gives an estimate of self-broadened resonance linewidth of
$\Gamma= 1 / T_2 \simeq 10$ GHz, which is in reasonable agreement
with the spectroscopic data on Rb pressure
broadening\cite{Rb_self_broadening_1979OptCom}. The maximal delays
measured in this experiment exceeded $30$ ps, corresponding to a
temporal dynamic range of about $1000$.

While the above experiments demonstrate the possibility of characterization of
ultrashort pulses with $10$ fs resolution with just a few picowatts of signal
power, the performance of the method is not limited by these values. The
maximal resolution and sensitivity achievable in this approach are determined
by the integration time, which, in turn, is limited by the stability of the
characterization setup and of the source itself. This potentially enables
complete characterization of ultrashort scattering from microscopic sources,
such as single molecules or nanoparticles, a regime relevant for coherent
nonlinear microspectroscopy
\cite{Nirit_Dan_SinglePulseCars_Nature02,CARS_microscopy_Xie_PRL99}.

The integration time necessary to achieve a given level of temporal resolution
can be reduced by using an array of fast detectors, either proportional or
single photon counting, for spectrally multiplexed measurements. Another
possible direction that can be explored is the characterization of ultrashort
pulses in the UV and XUV range. The accumulative nature of this method makes it
potentially suitable for characterization of attosecond pulse sources with
megahertz repetition
rates\cite{Plasmon_Array_High_Harmonic_Generation_2008Natur}.

In conclusion, we presented and demonstrated experimentally a self-referencing
ultrashort pulse characterization technique that combines the single-photon
sensitivity and versatility of linear detection with the temporal resolution
previously achieved only by nonlinear optical methods. This technique can
become a potent tool for ultrafast light sources characterization and for the
analysis of the fast dynamics of microscopic physical systems.

The authors would like to acknowledge financial support by the
Israeli Ministry of Science Tashtiyyot program and by the Crown
center of photonics. O.R. acknowledges support by the Converging
technologies fellowship of the Israeli Ministry of Science. O.S.
acknowledges support by the Adams fellowship program.

\end{document}


\title{\huge Supplementary Information \\ Shot noise limited characterization of femtosecond light pulses}

\author{O. Schwartz}
\email{osip.schwarz@weizmann.ac.il}
\author{O. Raz}
\author{O. Katz}
\author{N. Dudovich}
\author{D. Oron}
\affiliation{Department of Physics of Complex Systems, Weizmann Institute of
Science, Rehovot, Israel}

\maketitle

\section{Arrival time estimation}\label{processing}
The main vehicle of this work was extracting the arrival times of femtosecond
optical pulses from the electronic signal. A perfectly stable pulse would
produce a statistical distribution of electronically detected arrival times
with a probability density function $f(t)$ due to the jitter in the detector
and in the timing circuitry. This distribution is assumed to be constant
throughout the experiments up to a shift in time, which may depend on
temperature, pulse intensity and other variable parameters.  Such shifts do not
deteriorate the timing measurements since in all experiments described in this
work we estimate a relative delay between a pair of pulses, rather than an
absolute temporal position of a single pulse. The time correlated single photon
counting system (TCSPC) recorded the signal as a histogram of this distribution
with $4$ ps bins. A sample histogram of such distribution produced using a $30$
fs optical pulse is shown in Supplementary Fig.~\ref{histogram}. We also assume
that the small contribution of the femtosecond pulse shape to $f(t)$ is
constant throughout the measurement.
\begin{figure}
\includegraphics[width=90mm]{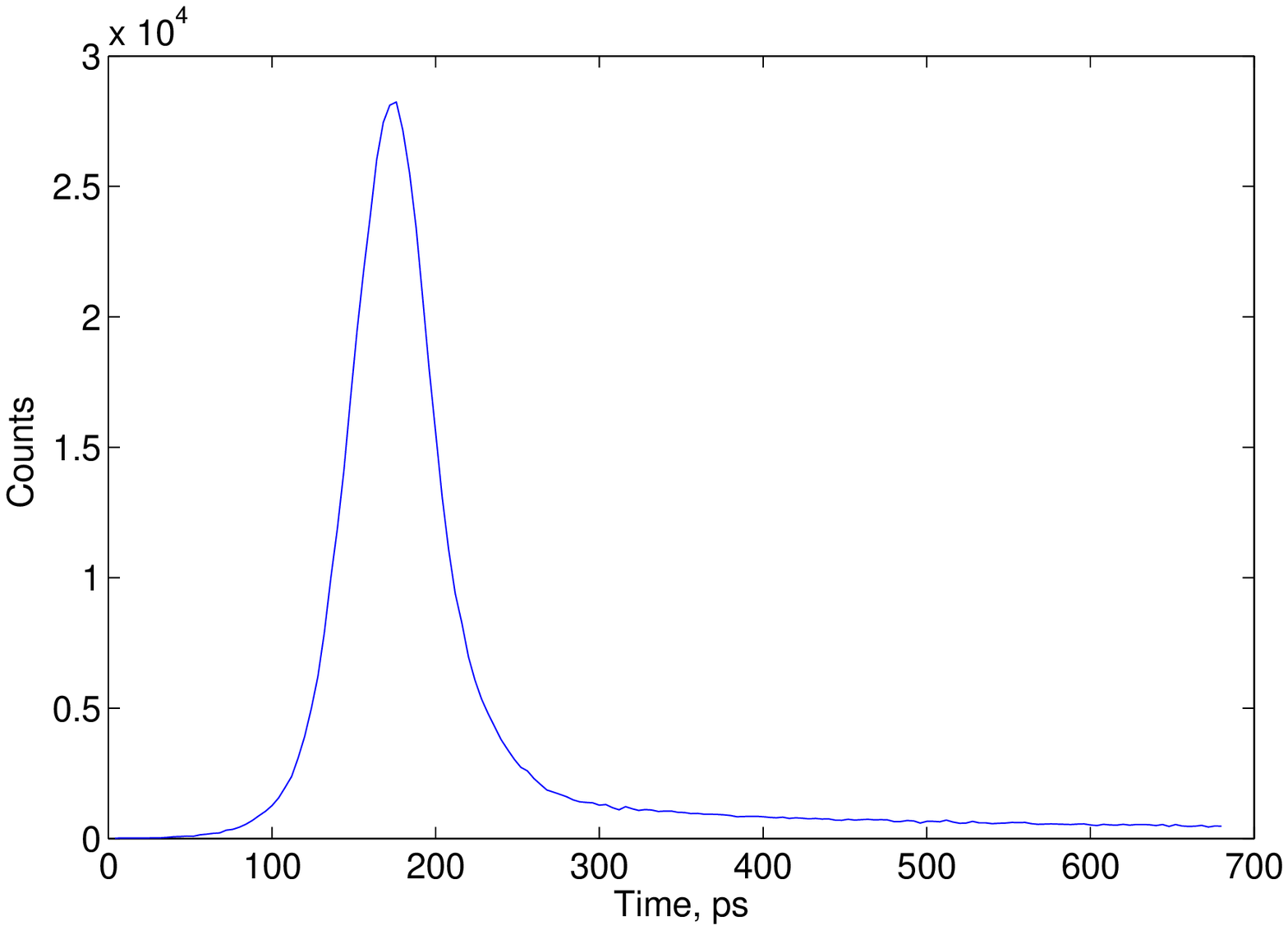}
\caption{A sample histogram obtained by directing an attenuated $80$ fs pulse
train into the detector, collection time $0.5$s}\label{histogram}
\end{figure}

The rigorous way of determining the arrival time with the least possible error
is to use the maximum likelihood estimation of the delay. The variance of such
estimate is determined by the Cramer-Rao bound, which in this case is given by
\begin{equation}\label{Rao}
\sigma_{\mbox{\tiny{min}}}^2 = \frac{{\Delta^2}}{N} \cdot
\left(\sum_i{\frac{\left(f(t_{i}+\Delta/2)-f(t_{i}-\Delta/2)\right)^2}{\int_{t_{i}-\Delta/2}^{t_{i}+\Delta/2}
f(t) dt}}\right)^{-1}
\end{equation}
where $\Delta$ is the bin width, $N$ is the number of detected photons, and
$t_i$ is the center of i-th bin in the histogram. For our system this estimate
yields $\sigma_{\mbox{\tiny{min}}}=39\mbox{ps}/\sqrt{N}$.

In this work, instead of using maximum likelihood estimation, we utilized
nearly as efficient but more economic and robust approach of multiplying the
histogram by a gaussian 'cap' and finding the centroid of the product. The
position of the cap was adjusted iteratively to coincide with the centroid.
This procedure is widely used for emitter localization in super-resolution
microscopy (see
ref. 17). 
The results proved to be practically insensitive to the RMS width of the cap in
the range $28$-$60$ ps; in this work, we used a gaussian cap with $32$ ps RMS
width. The efficiency of this approach can be assessed by comparing the
resulting measurement error $\sigma_{\mbox{\tiny{exp}}}$ to
$\sigma_{\mbox{\tiny{min}}}$. To compute the measurement error, we generated a
sample consisting of $5\times10^4$ estimates of a fixed relative delay between
two pulses (see Supplementary Section \ref{stability}), each estimate based on
data accumulated for $0.5$ s. The standard deviation within this sample was
found to be $\sigma_{1/2 s} = 68$ fs, leading to $\sigma_{\mbox{\tiny{exp}}} =
45\mbox{ps}/\sqrt{N} = 1.15\, \sigma_{\mbox{\tiny{min}}}$.

\section{Stability Analysis}\label{stability}
The $1/\sqrt{N}$ scaling of the temporal localization error holds as long as
the uncorrelated in time, or `white', timing jitter is the dominant source of
error. In most experimental systems there are, however, sources of errors that
increase with the integration time, such as flicker noise. Such noises
effectively limit the resolution improvement attainable by increasing the
integration time. The experiments presented here rely on $1/\sqrt{N}$ error
scaling to achieve femtosecond timing accuracy. It is therefore critical to
assess the noise sources in the system.

To this end, we carried out two experiments. In the first, two attenuated
replicas of a $30$fs pulse train, delayed with respect to each other by
approximately $3$ ns, were directed into the same single photon detector,
producing $1.8\times 10^6$ counts per second each. Histograms of arrival times
of the electric pulses from the detector were continuously recorded with $0.5$
s integration time. The relative delay between the two replicas was estimated
for every histogram. An Allan variance plot, showing the standard deviation of
the relative delay as a function of the integration time, is shown in Fig.1b in
the manuscript (blue line). The red line corresponds to the position of one of
the peaks without reference to the other. The relative delay data from this
experiment was used to estimate $\sigma_{\mbox{\tiny{exp}}}$ in Supplementary
Section \ref{processing}. The dashed black line represents the Cramer-Rao
bound, which determines the minimal error attainable in the absence of any
noise sources except `white' timing jitter.

The plot shows that under the conditions of this experiment, the error of the
relative measurement scales as $1/\sqrt{N}$ up to a thousand seconds
integration time, which corresponds to two femtosecond accuracy. At longer
integration times the error increases, which indicates the presence of a `red'
noise in the system. It can be attributed to some kind of creep in the optical
setup or in the electronics used. However, this noise becomes noticeable only
at integration time longer, or at accuracy significantly higher than that used
in the pulse characterization experiments presented here.

In the second experiment, we carried out a spectral delay
measurement of a transform limited pulse, focusing on the scaling of
the measurement error with the integration time. To gather the error
distribution statistics, the data were continuously collected in the
experimental conditions similar to those used in the spectral delay
measurements presented in Fig.2 and Fig.3 of the main text. An Allan
variance plot for each of the 20 slit positions, as a function of
the integration time per spectral slice, is shown in Supplementary
Fig.\ref{Allan_spectral}. The detection rate changes with the slit
position according to the spectrum of the pulse. The Cramer-Rao
bound, also shown in Fig.\ref{Allan_spectral}, is calculated for the
slit position producing the maximum detection rate (and thus the
lowest error). It can be seen that the localization of every slice
starts out as $\sqrt{N}$ at short integration times.  The onset of
the deviation from the $\sqrt{N}$ scaling occurs earlier than in the
experiment with a constant delay, due to increased duration of the
experiment required to reach a given integration time per spectral
slice. Nevertheless, the localization accuracy continues to improve
with the integration time during the entire experiment, reaching on
average $8.9$ fs standard deviation at integration time of $32$ s.
\begin{figure}
\includegraphics[width=90mm]{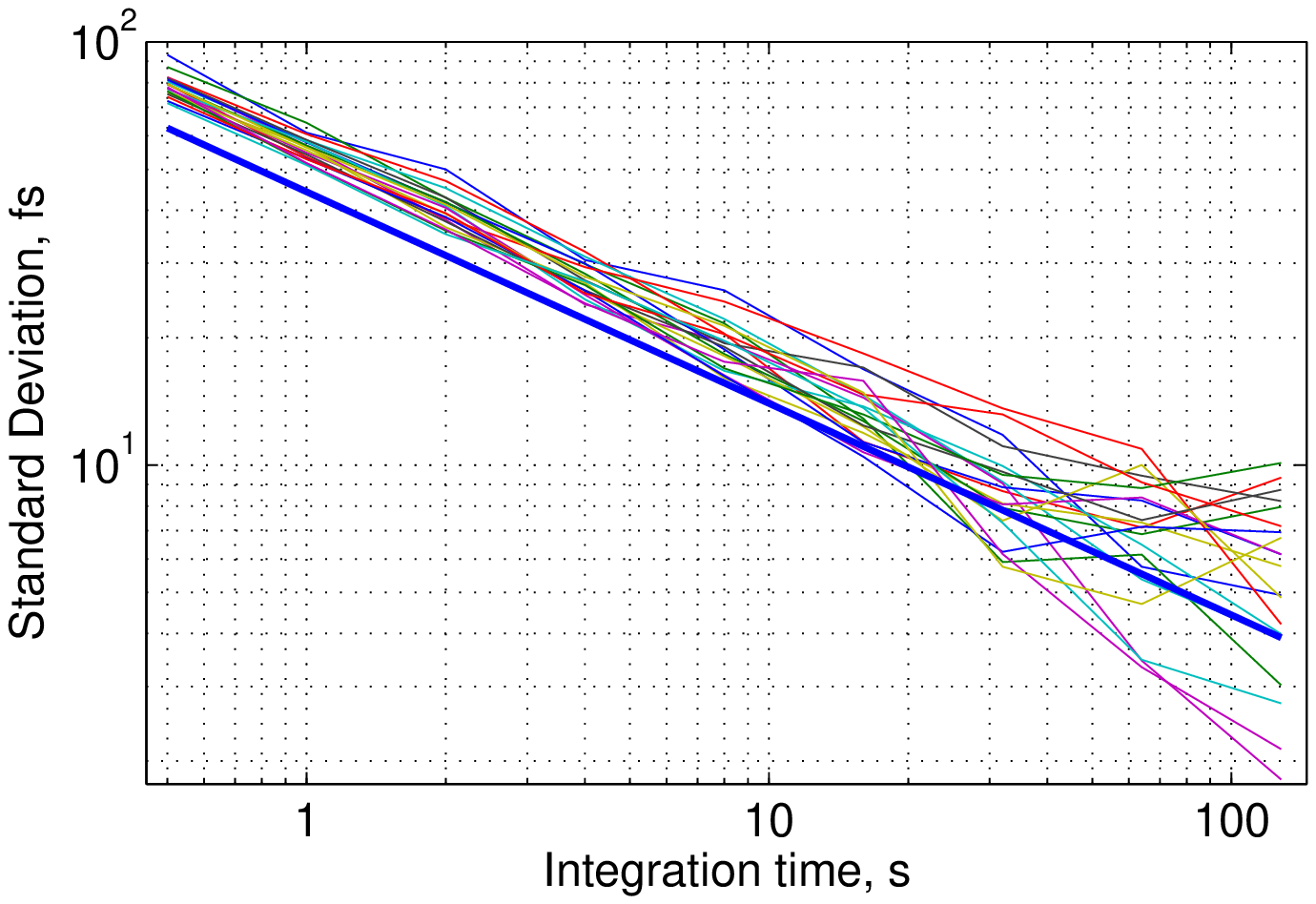}
\caption{\textbf{Allan variance plot for 20 slit positions.} Timing
error as a function of the integration time per spectral component
for 20 slit positions. The thick blue line corresponds to the
Cramer-Rao bound, calculated for the slit position corresponding to
the maximum of spectral intensity. }\label{Allan_spectral}
\end{figure}

\section{Time Domain Reconstruction}\label{time_domain}%
The reconstructed time-domain waveform of the pulse dispersed by
passing through a 6" slab of F3 glass (see Fig.~2)  is presented in
Supplementary Fig.~\ref{time_domain}.
\begin{figure}
\includegraphics[width=90mm]{time_domain_FROG_us.eps}
\caption{\textbf{Temporal profile of the pulse dispersed by 6" of F3 glass.}
Dashed lines represent phase, solid lines represent pulse intensity. Blue
lines: FROG measurements, red lines: our technique. }\label{time_domain}
\end{figure} The time-domain waveform, $E(t)$ was calculated by
Fourier-transforming the measured spectral-domain result
$E(\omega)=|E(\omega)|e^{i\phi(\omega)}$. The spectral delays $t(\omega)$,
weighted by the spectrum of the pulse, were fitted with a $3^{rd}$-order
polynomial $\tilde{t}(\omega)$. The spectral phase $\phi(\omega)$ was then
calculated by integrating this polynomial: $\phi(\omega)=\int^\omega
\tilde{t}(\omega')d\omega' $. The electric field spectral amplitude,
$|E(\omega)|$, was obtained by measuring the spectrum of the pulses with an
optical spectrum analyzer (ANDO AQ-6315A).

\section{Rubidium experiments}\label{Rubidium}
To verify the reconstruction of the pulses passing through the
Rubidium cell the measured spectral-delays and amplitudes were
fitted with the theoretical curves for a transmission through a
resonant medium. The transmitted spectral amplitudes were obtained
from the spectrum recorded with 0.05 nm resolution by an optical
spectrum analyzer. Rubidium has two resonant lines in the relevant
spectral range, each resonance modifies the electric field of the
pulse by multiplying it with a factor of $f(\omega) =\exp\left[ -
\alpha l \;\left(1\left/(1 + i\omega T_2)\right.\right)\right]$,
where $T_2$ is the inverse pressure broadened linewidth, $\alpha l$
is the optical density, and $\omega$ is the frequency detuning (see
ref. 23). The corresponding spectral delay is then $t(\omega) =
\alpha l T_2 \frac{1- \omega^2 T_2^2}{\left( 1 + \omega^2
T_2^2\right)^2}$. The width of the Rb absorbtion lines is dictated
by the very large optical density of the vapor cell and is much
larger than the natural linewidth. However, even these broadened
lines are too sharp to be resolved with the spectral resolution of
our pulse shaper, which was set to 0.5 nm ($\simeq 250GHz$).
Nonetheless, the tails of the peaks in the spectral delay curve,
starting from detuning of ~1nm, are well resolved. Fitting
simultaneously the measured spectral delays of these Lorentzian
tails and the measured spectral transmitted amplitudes, with the
theoretical formulas presented above, allows for the extraction of
the two fit parameters, $T_2$ and $\alpha l$.  The measured spectral
amplitudes were fitted in the entire spectral range to the
theoretical formula convolved with a 0.05nm FWHM Gaussian instrument
spectral response. The extracted values for these parameters for the
210\textcelsius\ measurement are $1 / T_2 = 10$GHz and $\alpha l =
1400$, which is in reasonable agreement with the spectroscopic data
on Rb pressure broadening.